\documentclass{aa}
\usepackage{graphicx}
\begin{document}

\title {Did VV~29 collide with a dark Dark-Matter halo?}

\author{F.H. Briggs\inst{1}, O. M\"oller\inst{1}, J.L. Higdon\inst{1,2},
N. Trentham\inst{3}, E. Ramirez-Ruiz\inst{3}}
\institute{Kapteyn Astronomical Institute, P.O. Box 800, 9700 AV
  Groningen, The Netherlands\\
\and Department of Astronomy, Cornell University, Ithaca, NY 14853, U.S.A. \\
\and Institute of Astronomy, Madingley Road, Cambridge CB3 0HA, United Kingdom
}

\date{Received ~; accepted ~}

\titlerunning{VV~29}
\abstract{
  Westerbork Radio Synthesis Telescope observation of the
galaxy VV~29$=$Arp~188$=$UGC~10214 shows that there are at least three
distinct dynamical components whose kinematics can be traced
in 21cm line emission.  The system appears to be the result of a
galaxy-galaxy interaction.   We identify  a sufficient
number of dynamical elements containing baryons (stars and
neutral gas) that there is no compelling reason
to postulate the presence of an additional dark matter halo that
is devoid of detectable baryons.\\
\indent ~ ~ ~The central
galaxy VV~29a is massive ($V_{rot}\sim 330$~km~s$^{-1}$) and gas
rich ($M_{HI}\sim 6 {\times}10^{9}h^{-2}{\rm M}_{\odot}$). The distinctive
optical plume (VV~29b), which extends eastward from the main galaxy, is also
gas rich ($M_{HI}\sim 3{\times}10^{9}h^{-2}{\rm M}_{\odot}$) and has a very
low gradient in line of sight velocity ($<30$~km~s$^{-1}$) over
${\sim}70h^{-1}$kpc. On  the western side, there is an HI feature of
$M_{HI}\sim 4{\times}10^{8}h^{-2}{\rm M}_{\odot}$
 that participates strongly in orbital motion
about the host in the same sense of rotation as the VV~29a itself.
A blue, less massive, gas-rich galaxy ``VV~29c''
($M_{HI}\sim 9{\times}10^{8}h^{-2}{\rm M}_{\odot}$)
appears clearly in the HI maps as an $\sim$170~km~s$^{-1}$ wide
spectral feature,
seen in projection against or, more likely, behind the west side of the host disk.
Its high recessional velocity is counter to the host rotation direction.
The optical images of Trentham et al (2001) show signs of this
blue dwarf against the redder VV~29a disk.\\
\indent ~ ~ ~The companion galaxy CGCG~27-021$=$MGC~09-26-54
(at projected distance ${\sim}115h^{-1}$kpc) is not detected
in 21cm line emission ($M_{HI}<10^{9}h^{-2}$M$_{\odot}$). \\
}

\maketitle

\section{Introduction}

The interacting system VV~29 (CGCG~275-023=Arp~188=UGC10214)
has attracted recent attention for possibly being the victim
of a collision with a pure dark matter halo (Trentham,
M\"oller \& Ramirez-Ruiz 2001). This paper describes Westerbork
Synthesis Radio Telescope observations that measure the
neutral gas kinematics and show that there are numerous
dynamical components in this interacting group that can be
detected through the electromagnetic emission from their
baryons.

The system is a member of several historical catalogs.
The Atlas and Catalog of Interacting Galaxies
(Vorontsov-Velyaminov 1959) listed the system  as VV~29 and
additionally assigned the label VV~29a to the
bright spiral galaxy at the center of the system, as well as
calling the faint optical feature to the east ``VV~29b,''
as though it were a galaxy in its own right.  Arp included
the system in his
Catalog of Peculiar Galaxies (1962, 1966), where the morphology of
the thin optical plume VV~29b motivated its placement in
the category of ``Narrow Filaments.''  Arp added
a comment that there is a disturbance inside the western spiral
arm of the host (VV~29a) and that ``the filament may originate
there.''   The new WSRT observation identifies this disturbance
with a separate galaxy whose velocity is distinct from the host
VV~29a and plume VV~29b components. We name the newly identified
galaxy VV~29c.  These three components are labeled on the images
in Figs.~\ref{NHI_GALAXY.fig} and \ref{NHI_DWARF.fig}.

A Nan\c{c}ay Radio Telescope detection of VV~29 provided a redshift of
9401$\pm$15~km~s$^{-1}$ (Bottinelli et al 1993),
but the WSRT observations show that the
narrow  emission feature in the Nan\c{c}ay spectrum
emanates from the tidal plume component VV~29b.
The new observation shows that the
host galaxy VV~29a has a redshift $\sim$100~km~s$^{-1}$
less.

VV~29 finds itself a member of the poor cluster WBL~608
(White et al 1999).  Joseph et al (1984) paired VV~29 with
CGCG~27-021 $=$MGC~09-26-54,
an early type system with heliocentric redshift of
9439 $\pm$ 27 km~s$^{-1}$ located at a projected distance
of 115$h^{-1}$~kpc (where is $h$ is the normalized Hubble
parameter $h=H_o/100$~km~s$^{-1}$Mpc$^{-1}$).

The generation of long filamentary tails, counter-tails,
and bridges in galaxies through tidal forces is well understood,
thanks to concerted observational and theoretical efforts over the past
three decades (for example, Toomre $\&$ Toomre 1972, Schombert et al. 1990).
In this view, VV~29 represents an ongoing merger between
a massive spiral and one or more smaller companions, including
one that may be partially
obscured or disrupted as a result of the interaction.
Similar interacting systems are common in the Arp Catalog.
Blue tidal features such as the VV~29b plume are generally HI rich
(Higdon \& Wallin 2001), making aperture synthesis observations in
the 21cm line extremely valuable in unraveling the interaction
history of peculiar systems, and in revealing the presence of obscured or
disrupted companion galaxies through their distinct kinematic components.
For example, Hibbard \& van Gorkom (1996) presented extensive
optical and HI observations of a merging sequence of galaxies
(Toomre 1977)
whose properties resemble the VV~29 system that we study here.

\section{The WSRT Observations}
 
The Westerbork Synthesis Radio Telescope observed VV~29$=$UGC~10214
at celestial coordinates
16$^{\rm h}$06$^{\rm m}$03$^{\rm s}$.9, 55$^{\circ}$25$'$32$''$(J2000)
for 10 hours on 15-16 April 2001. Calibration scans on 3C286 for
64 minutes preceding VV~29 and 3C48 for 23 minutes following VV~29
provided complex gain and passband calibrations. An adopted flux
density of 14.94 Jy for 3C286 sets the flux scale.

The spectrometer recorded 128 frequency channels in two linear
polarizations to cover a 5 MHz band centered on
$z=0.031355$ in the Heliocentric reference frame. This provides
a channel spacing of 8.5~km~s$^{-1}$. The first five channels (lowest redshift)
and last eight
channels showed deteriorating response, since they are close to the
band edge; deleting them leaves a useful band covering a
velocity range of $\sim$980~km~s$^{-1}$.

Application of  standard calibration, mapping, and
deconvolution steps in the Classic
AIPS aperture synthesis imaging system
led to a data cube of channel maps with
synthesized beam resolution of $36''{\times}26''$
(major axis at position angle $8^{\circ}$) with a ROBUST parameter
of 3 in order to achieve an optimal sensitivity to weak emission.
An unresolved continuum source ($300$~mJy) at
16$^{\rm h}$06$^{\rm m}$07$^{\rm s}$.6, 55$^{\circ}$21$'$35$''$(J2000)
 to the south of the target coordinates
provides a secondary flux density calibration check
as well as permitting
refinement of the complex gain calibration through the course of
the observation
using self-calibration procedures. Model visibilities for
the bright continuum source
were subtracted from the UV dataset prior to construction of
the cube of channel maps. Further continuum subtraction using
linear interpolation between the extreme channels that were free
of line emission was performed once the cube of channel maps had
been computed. The surprisingly wide velocity spread of the HI
associated with this system meant that there were few emission-free
channels, and this adds some additional, systematic noise to the
channel maps.  The rms noise fluctuation in emission-free regions of
the final channel maps was typically $\sim$0.33~mJy per beam.

\section{Analysis}

Close inspection of the individual channel maps and
cuts in distance-velocity planes revealed the nature of
the principal HI features: 1. the host galaxy VV~29a itself,
2. a second
galaxy of lower HI mass that is viewed along the same line of sight
as the western side of the host, and 3. the eastern plume (detected
in HI coincident
with the  distinctive optical plume). A possible counter-plume
can be identified through the presence of
faint HI emission on western side of the host. This section describes the
observational evidence.

\begin{figure}
  \includegraphics[width=8.5cm]{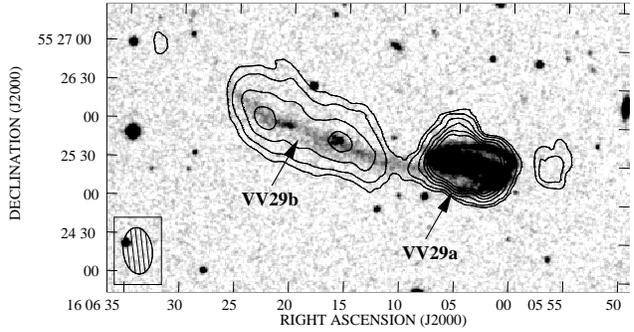}
  \centering
  \caption[]{VV~29a and VV~29b: Integral HI column density contours for the velocity
range $-400$ to $+240$~km~s$^{-1}$ relative to $z=0.031355$ are superimposed
on a DSS image at high contrast. This
range includes the full velocity spread of the host galaxy VV~29a and the
two plume components, while excluding the signal from the blue
dwarf. The synthesized beam shape is shown at the lower
left.  Contour levels are
0.6, 1.2, 2.4, 3.6, 4.8, 5.9 ${\times}10^{20}$ H atoms
cm$^{-2}$.}
  \label{NHI_GALAXY.fig}
\end{figure}

\begin{figure}
  \centering
  \includegraphics[width=8.5cm]{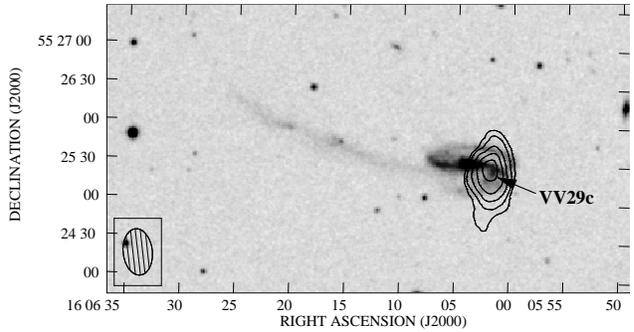}
  \caption[]{VV~29c: Integral HI column density contours for the velocity
range $+240$ to $+470$~km~s$^{-1}$ to accept the full range of
the HI signal of the blue dwarf VV~29c are superimposed on the DSS image
at low contrast.
Contour levels are 0.6, 1.2, 2.4, 3.6, 4.8 ${\times}10^{20}$ H atoms
cm$^{-2}$.}
  \label{NHI_DWARF.fig}
\end{figure}

\begin{figure}
  \centering
  \includegraphics[width=8.5cm]{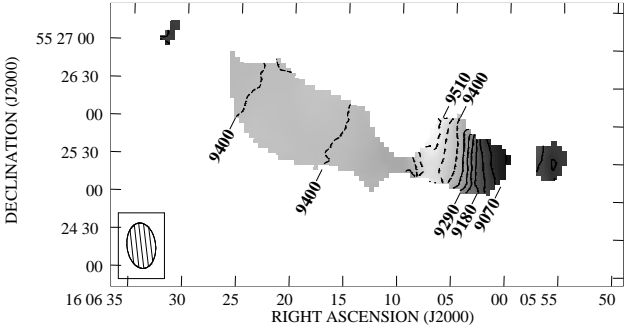}
  \caption[]{VV~29a and VV~29b: Velocity field computed from the first moment of
the emission profiles for the gas in the velocity
range $-400$ to $+240$~km~s$^{-1}$ relative to $z=0.031355$.
Contour levels are spaced at 55~km~s$^{-1}$ intervals.}
  \label{VELF_GALAXY.fig}
\end{figure}

\begin{figure}
  \centering
  \includegraphics[width=8.5cm]{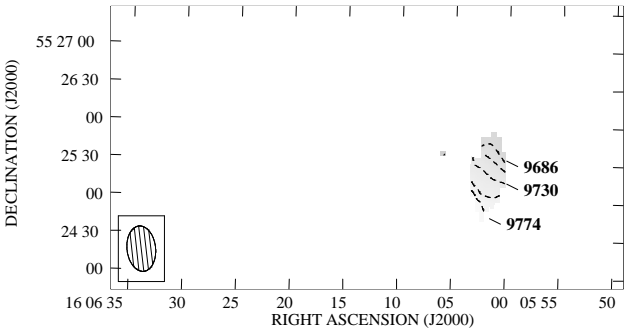}
  \caption[]{VV~29c: Velocity field computed from the first moment of
the emission profiles for the gas in the velocity
range $+240$ to $+470$~km~s$^{-1}$ relative to $z=0.031355$. Lighter
shading represents higher redshifted velocity.
Contour levels are spaced at 22~km~s$^{-1}$ intervals. }
  \label{VELF_DWARF.fig}
\end{figure}

\begin{figure}
  \centering
  \includegraphics[bb= 118 144 492 718,width=8.5cm,clip]{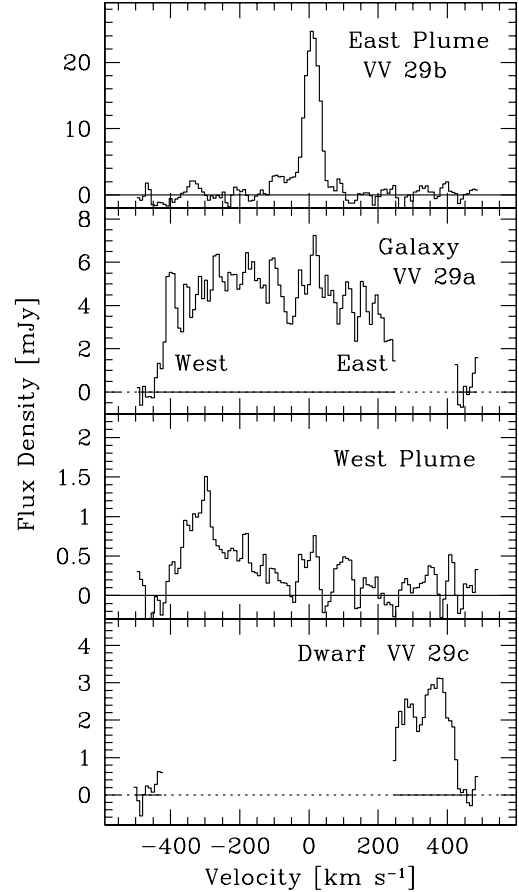}
  \caption[]{Integral spectra for the four principal kinematic
components obtained by summing pixels in the channel maps
in the four regions. The zero velocity corresponds to $z=0.031355$
(Heliocentric).}
  \label{spectra.fig}
\end{figure}

Figures~\ref{NHI_GALAXY.fig} and \ref{NHI_DWARF.fig} show contours of
the integral HI emission overlaid on DSS images in order to show the
correspondence  between the extended HI emission and the optical plume
and the location of the second gas-rich galaxy,  VV~29c.
Figures~\ref{VELF_GALAXY.fig} and \ref{VELF_DWARF.fig} show velocity
contours corresponding to the Figs.~\ref{NHI_GALAXY.fig}
and \ref{NHI_DWARF.fig}, respectively.  The velocity fields were obtained
from the first moments of the HI profiles, after subjecting the images
to automated thresholding computed from a spatially and spectrally
smoothed version of the datacube. Figure~\ref{spectra.fig} shows the
integral spectral profiles for the components that can be seen in the
maps.

Figures~\ref{NHI_GALAXY.fig} to \ref{spectra.fig} make it clear that
the galaxy itself is a large rotating disk system with a velocity
spread of $\sim$700~km~s$^{-1}$; the western side of the galaxy
has a projected velocity of approach along the line of sight,
while the eastern side recedes.
The systemic velocity falls at 9300~km~s$^{-1}$ (Heliocentric), which is
 $\sim$100~km~s$^{-1}$ less than the nominal redshift.
There is a significant distortion of the $N_{HI}$ contours extending
northward at the eastern side of the galaxy. This distortion is
coincident with very faint optical light in the deep images of
Trentham et al (2001) and is another symptom of severe
gravitational interaction.  A more detailed discussion of the features
in the optical light follows in \S~\ref{discussion.sect}.
The newly identified galaxy VV~29c has a systemic velocity
of 9730~km~s$^{-1}$, and thus its motion would appear to carry it counter to the
rotation direction of the host galaxy with a projected relative
velocity of $\sim$430~km~s$^{-1}$.  The rotational axis of VV~29c
is oriented roughly perpendicular to the rotational axis of VV~29a.

Re-examination of the optical imaging from Trentham et al (2001)
gives an estimate of the ratio of optical $u$-band light for
VV~29c relative to VV~29a of 1:6, which is comparable to the
ratio of approximately 1:7 in neutral hydrogen mass.

\begin{figure}
  \centering
   \includegraphics[width=8.5cm]{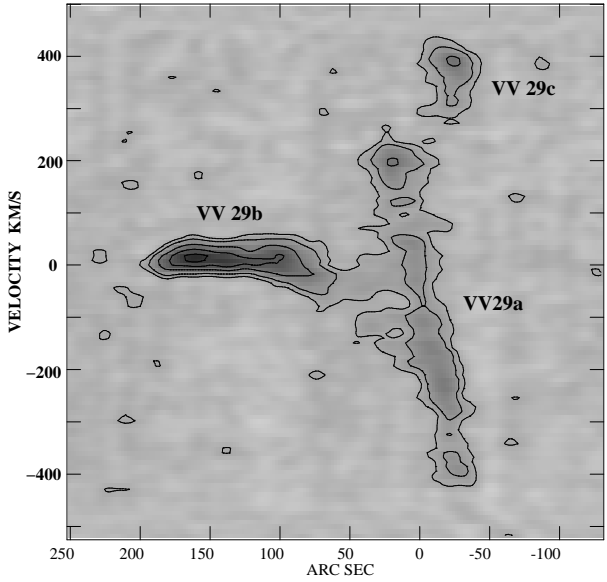}
  \caption[]{Contours of HI brightness temperature in the distance-velocity
plane along a line aligned with the eastern plume (at an angle
of 18.5$^{\circ}$ relative to a line of constant declination,
corresponding to position angle 71.5$^{\circ}$).
The zero of velocity is $z=0.031355$, and the origin of distance
falls at 16$^h$06$^m$04$^s$.8, 55$^{\circ}$25$'$10$''$(J2000).
Contour levels are 1, 2, 4, 6, 8 K.}
  \label{LV_PLUME.fig}
\end{figure}
\begin{figure}
  \centering
  \includegraphics[width=8.5cm]{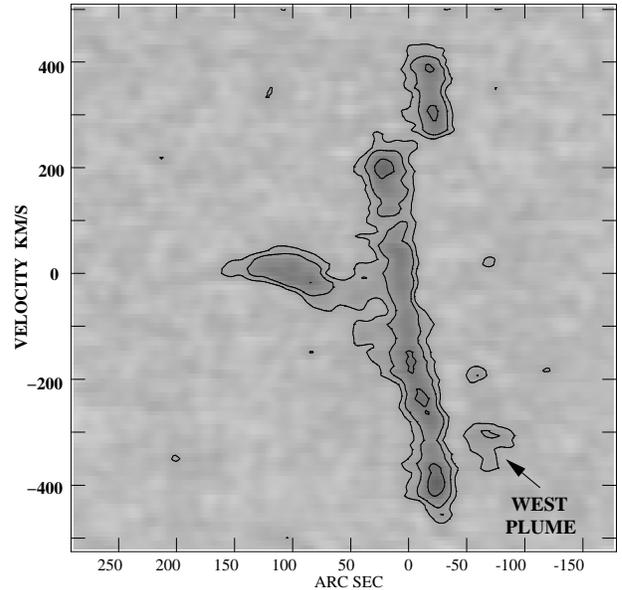}
  \caption[]{Contours of HI brightness temperature in the distance-velocity
plane along a line aligned with the major axis of the
host galaxy VV~29a at position angle $90^{\circ}$.
The zero of velocity is $z=0.031355$, and the origin of distance
falls at galaxy center at 16$^h$06$^m$03$^s$.9, 55$^{\circ}$25$'$32$''$(J2000).}
  \label{RA_V_MAJ.fig}
\end{figure}

The plume
to the east has very little velocity gradient along its length, falling
at $\sim$100~km~s$^{-1}$ higher redshift than the systemic velocity of
the host, and only near the base does the plume's velocity deviate toward
the systemic velocity of the host. A clearer picture of the velocity
structure  comes from Fig.~\ref{LV_PLUME.fig},
the distance-velocity plane contour map oriented along the main
body of the eastern plume. In this plot, the horizontal feature to
the left is the constant velocity plume. The steeply sloped linear
feature centered at position ${\sim}0''$, -100~km~s$^{-1}$ is the
host VV~29a, and the isolated feature to the upper right is a
cut through VV~29c.  The diffuse emission between the inner end
of the plume and VV~29a represents a dispersed extension of the
plume, which passes through zero relative velocity a little to the
east of the center of VV~29a, and then picks up velocity in the
sense of rotation of the host as it is traced further west. This
could signify that the plume connects to the disk, possibly to
one of the well defined spiral arms in the optical images.
Fig.~\ref{RA_V_MAJ.fig} presents a distance-velocity cut that is aligned
in an east-west direction along the major axis of the host, so that
it cuts through the inner end of the long eastern plume
and also the blob of emission
seen on the western side of VV~29a in Fig.~\ref{NHI_GALAXY.fig}. One
can smoothly extrapolate the rising speed of the inner edge of the
plume to the western blob, which in turn has a projected velocity
component along the line of sight in the
sense of rotation of the host; however, we discuss later (in
Sect.~\ref{discussion.sect})  the
morphology of the faint optical emission from the western side
of VV~29a, which leads us to associate the western blob with
a counter-tail that is distinct from the VV~29b plume.

\begin{figure}
  \centering
  \includegraphics[width=8.5cm]{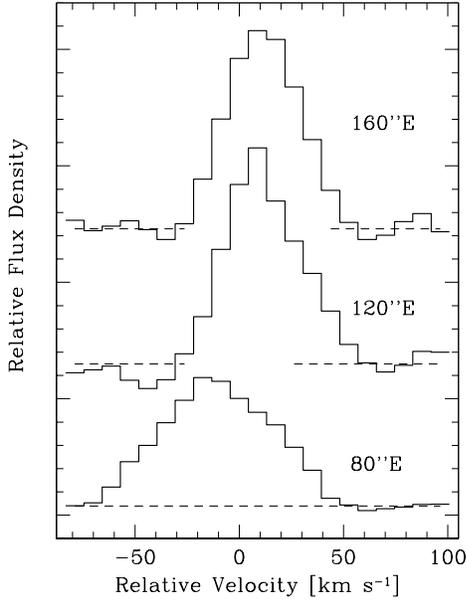}
  \caption[]{Spectra at three locations along the plume. The spectral
channel resolution is indicated by the histogram plot. Locations
are labeled in arcsecs eastward from the Right Ascension of the host
galaxy center.  Velocities are measured with respect to $z=0.031355$.
Baselines at zero flux density are indicated for each spectrum by
dashed lines. }
  \label{plume_spectra.fig}
\end{figure}

\begin{figure}
  \centering
  \includegraphics[bb=98 144 492 718,width=7.5cm,clip]{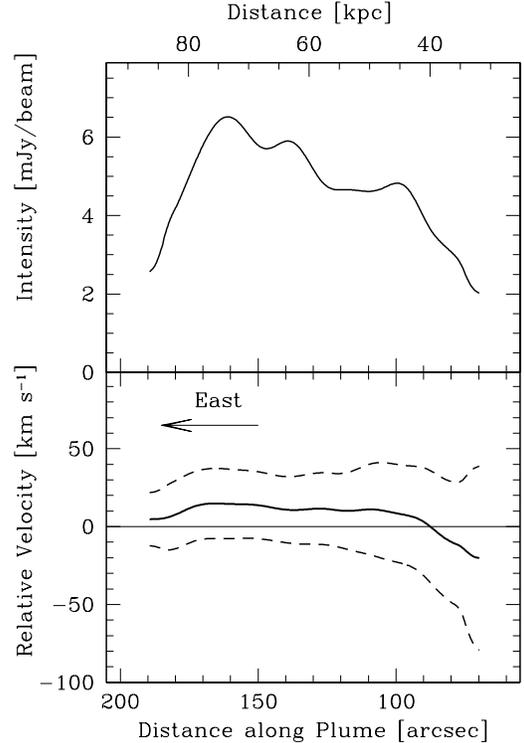}
  \caption[]{ Intensity, velocity centroid and velocity width from
fitted Gaussian profiles along the plume.  Velocity centroid (solid curve
in lower panel) is indicated with respect to $z=0.031355$, while velocity
spread(measured at FWHM) is indicated by the dashed curves in the lower
panel. A distance scale is drawn at the top assuming for Hubble Constant
$H_o=100$~km~s$^{-1}$Mpc$^{-1}$.}
  \label{plume_v_width.fig}
\end{figure}

Figures~\ref{plume_spectra.fig} and \ref{plume_v_width.fig}
explore the degree of velocity collimation along the main body of
the plume. Fig.~\ref{plume_spectra.fig} shows spectra at three
locations in the bright regions of the plume.  Fig.~\ref{plume_v_width.fig}
shows the peak intensity, velocity centroid and velocity width (FWHM)
of Gaussian fits to the spectra as a function of distance along the
plume. The velocity appears constant within 10~km~s$^{-1}$ over a distance
of $\sim$40$h^{-1}$~kpc along the plume.  The velocity width $\sim$40~km~s$^{-1}$
($\sigma\approx 17$ km~s$^{-1}$ of the profiles is relatively wide in
comparison to the velocity
dispersions $\sim$6-10~km~s$^{-1}$ that characterize the HI in disk
galaxies (van der Kruit \& Shostak 1982, 1984, Dickey, Hanson \&
Helou 1990, van Zee \& Bryant 1999).

\begin{figure}
  \centering
  \includegraphics[width=8.5cm]{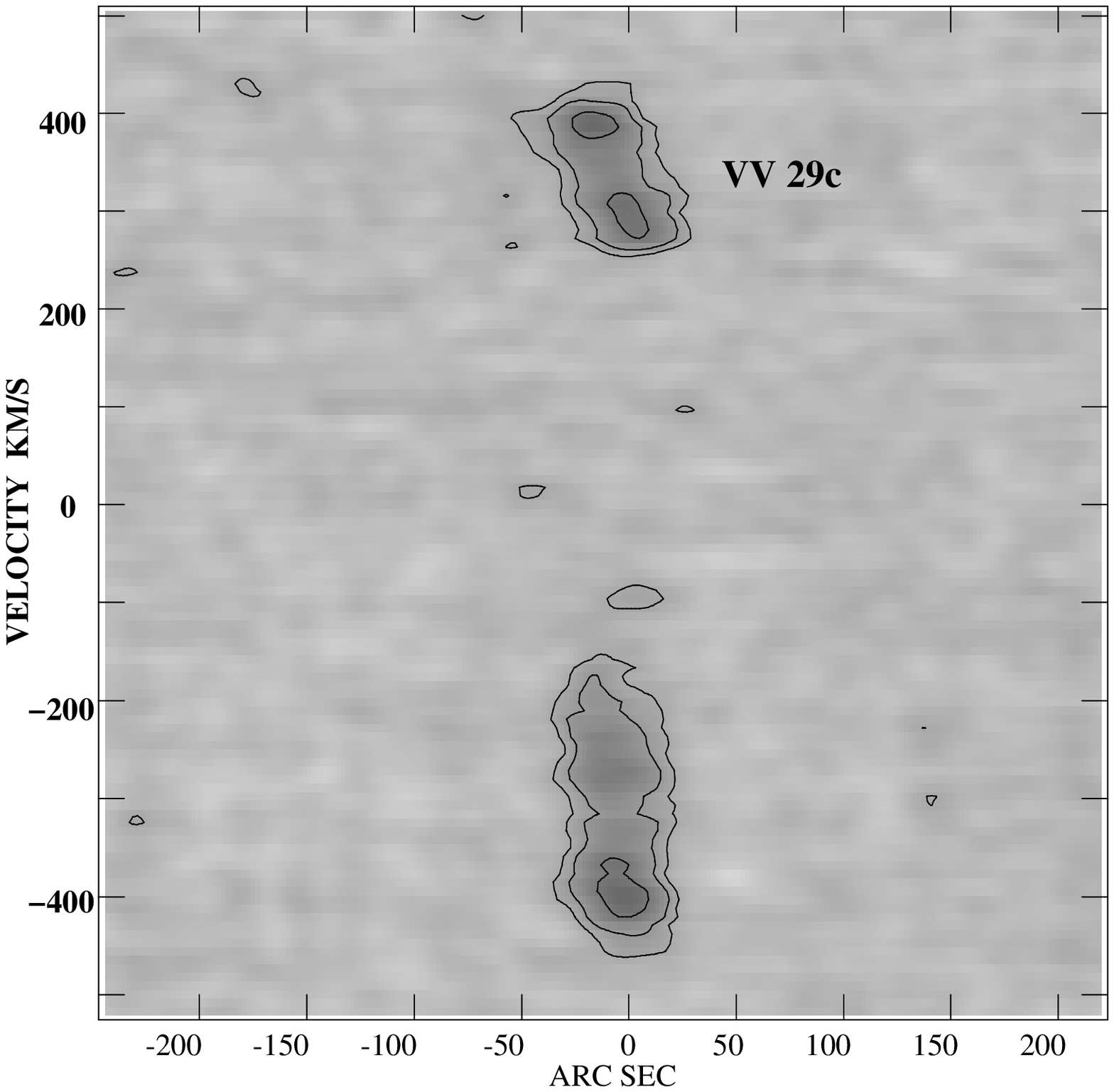}
  \caption[]{Contours of HI brightness temperature in the distance-velocity
plane along a line passing through the center of the blue dwarf VV~29c.
The distance axis is a line of constant right ascension.
The HI signal of the blue dwarf VV~29c falls in the upper region of the
plot in the velocity range 250 to 420~km~s$^{-1}$.
The zero of velocity is $z=0.031355$, and the origin of distance
falls at 16$^h$06$^m$02$^s$.0, 55$^{\circ}$25$'$32$''$(J2000).
Contour levels are 1, 2, 4 K.}
  \label{DEC_V_DWARF.fig}
\end{figure}

A final distance-velocity plane cut in Fig.~\ref{DEC_V_DWARF.fig}
along the major axis of the dwarf galaxy VV~29c
(perpendicular to the major axis of the host VV~29a) shows the distinct
velocity separation between the corotating gas in the disk on
the west side of the host VV~29a (blue shifted) and the redshifted
signal from VV~29c.  The signature of rotation of VV~29c
is also clear in this L-V plane.

\begin{figure}
  \centering
  \includegraphics[bb= 118 144 492 718,width=8.5cm,clip]{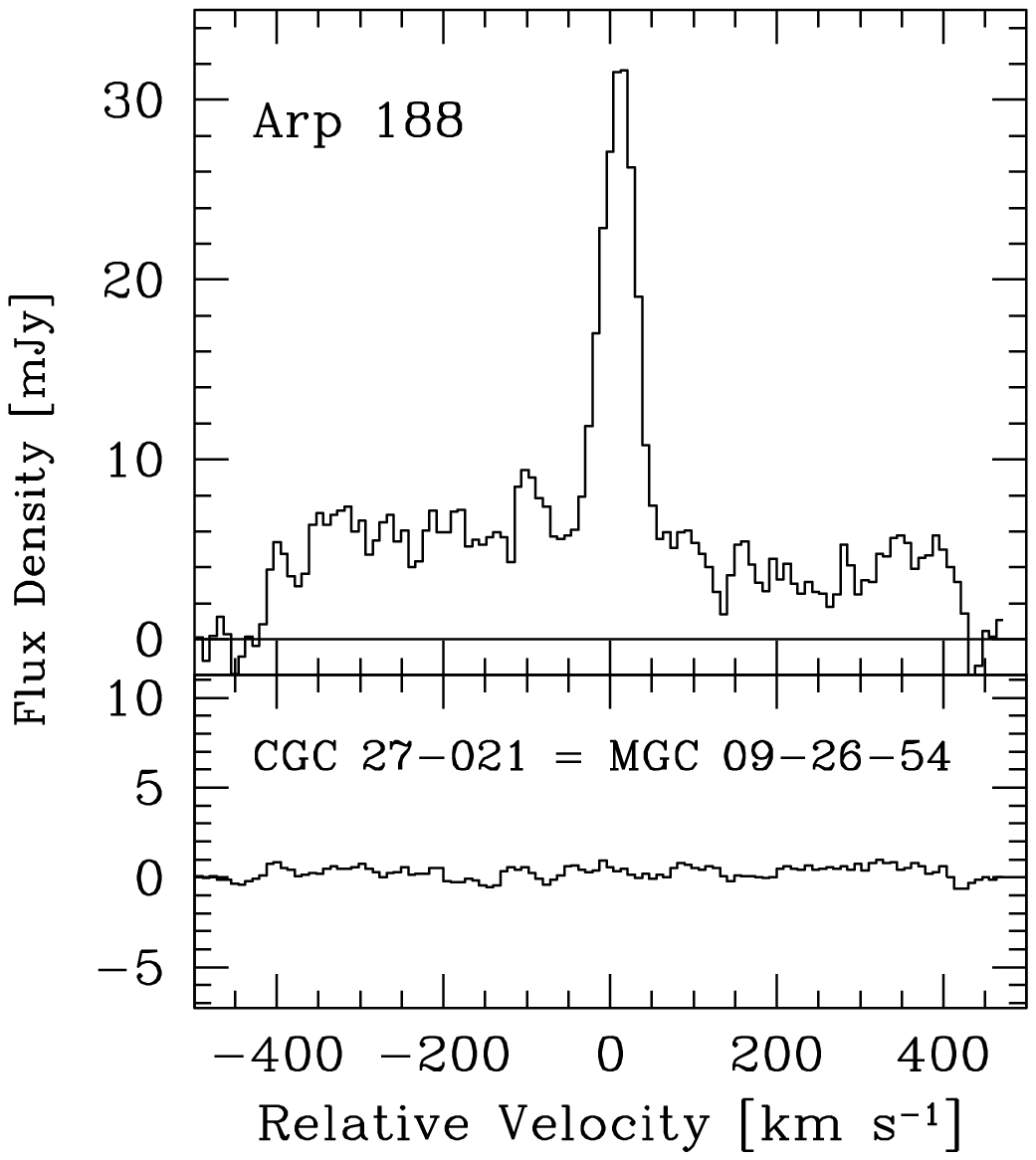}
  \caption[]{Integrated flux density spectra for VV~29$=$Arp~188$=$UGC~10214 and
MCG~09-26-54. The integral for the VV~29 system (top panel)
comes from summing the four spectra of Fig.~\ref{spectra.fig}.
The MCG~09-26-54 spectrum (lower panel) comes from integrating the
spectrum over the $52''{\times}52''$ patch (30~kpc$\times$30kpc)
centered on $16^{\rm h}05^{\rm m}48.^{\rm s}$, $55^{\circ}21'53''$
(J2000), the celestial coordinates of MCG~09-26-54.
The zero velocity corresponds to $z=0.031355$ (Heliocentric).
}
  \label{integral_spectra.fig}
\end{figure}

Figure~\ref{integral_spectra.fig} has plots of the integrated
spectrum (`single-dish') of the VV~29 system and the spectrum
of the region surrounding the early-type companion galaxy
MCG~09-26-54. The very broad component emitted by the host
galaxy is easy to miss in single-dish observations, as
evidenced by the Nan\c{c}ay spectrum (Bottinelli et al 1993),
which recovered only the narrow profile of the plume.
Figure~\ref{integral_spectra.fig}  shows no evidence of any
emission from MCG~09-26-54, implying an upper limit on its
HI mass
$(\Delta V/50$~km~s$^{-1})10^8h^{-2}\rm{M}_{\odot}$
that is directly proportional to the assumed velocity
width $\Delta V$.

\section{Discussion}
\label{discussion.sect}

\begin{figure}
  \centering
  \includegraphics[width=7.5cm,clip,angle=-90]{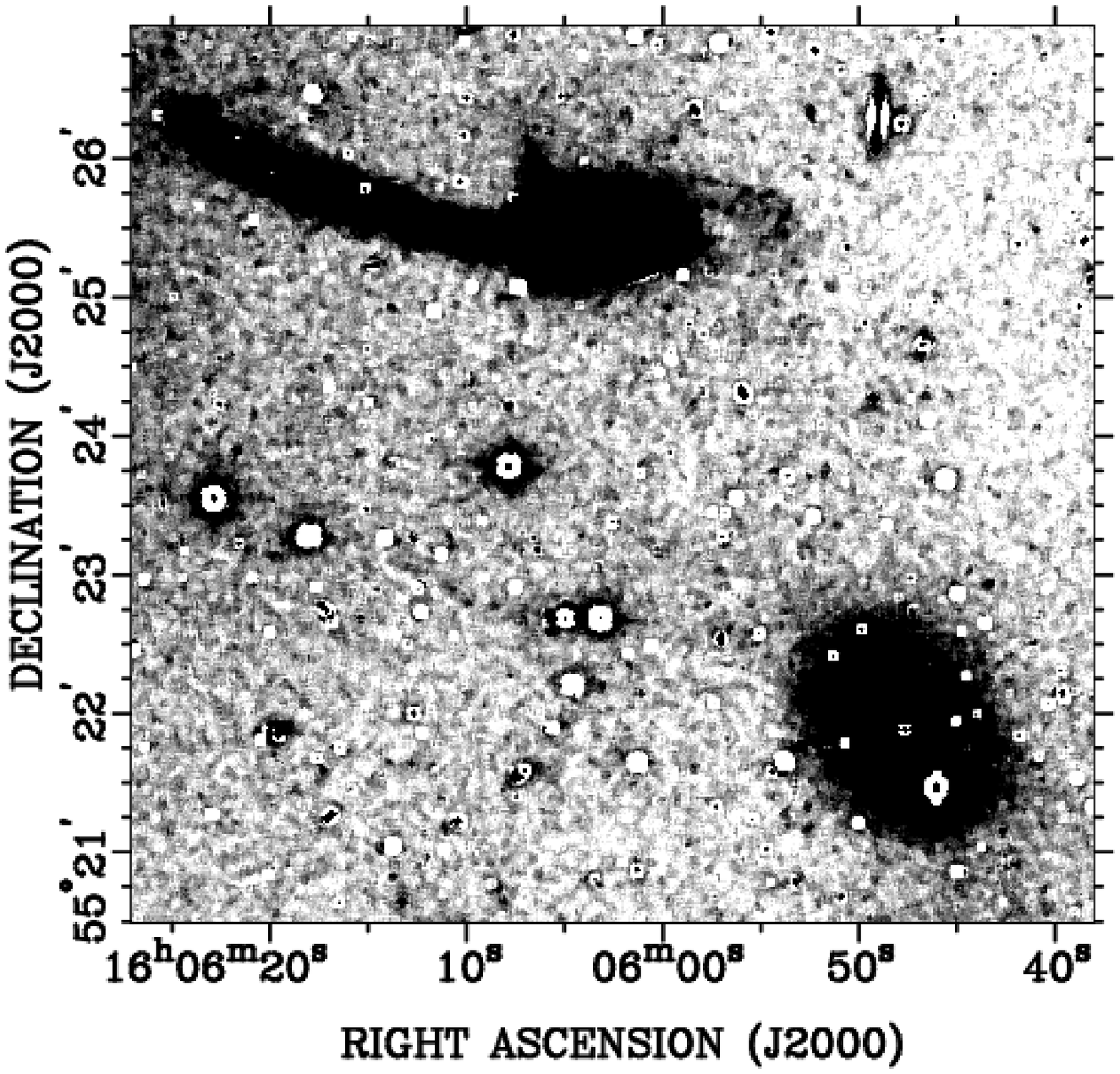}
  \caption[]{Smoothed, filtered, high contrast $g$-band image of
the VV~29 field, including the elliptical MCG~09-26-54 at lower
right.
}
  \label{deep_opt.fig}
\end{figure}

\begin{figure}
  \centering
  \includegraphics[width=8.5cm,clip]{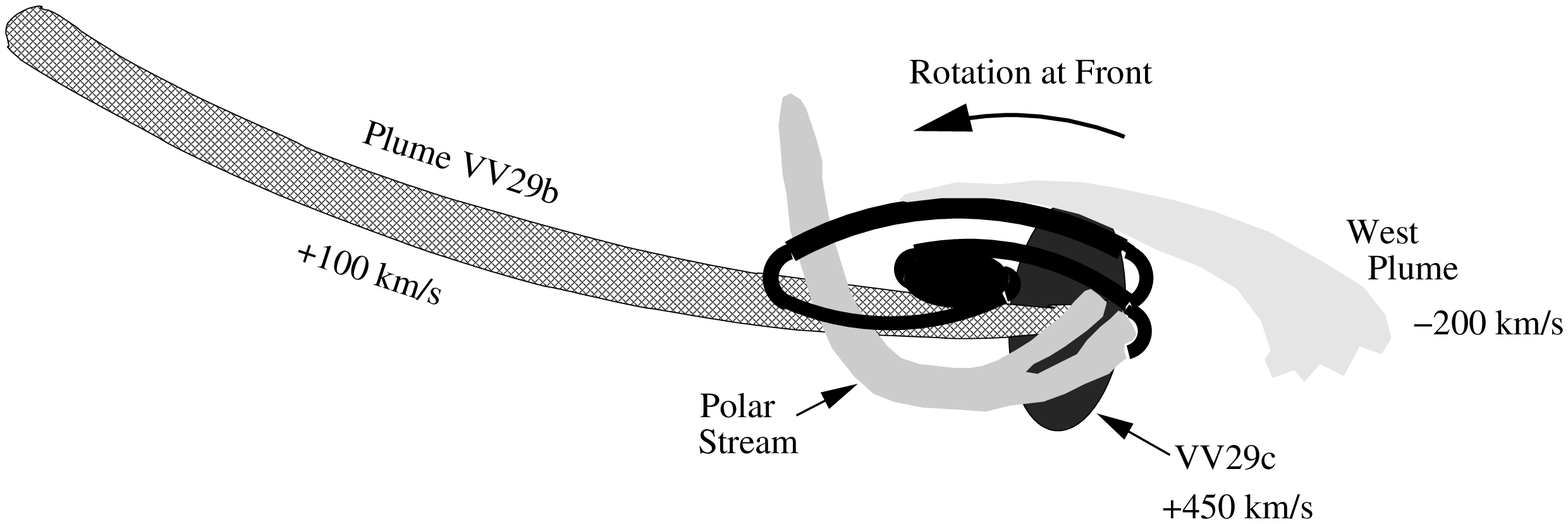}
  \caption[]{Inventory of structures in the VV~29 system.
Components are labeled with approximate line of sight velocities
with respect to the systemic velocity of VV~29a. The eastern
plume VV~29b, which passes behind the host galaxy VV~29a,
may connect to the `upper' spiral arm of on the western side
of VV~29a. The `polar stream' appears to  start on the `lower'
spiral arm and end by flipping up behind the eastern side of VV29~a,
where it the distorts the HI contours in Fig~\ref{NHI_GALAXY.fig}.
The western plume is detected in HI and faintly visible in
optical light. The blue dwarf galaxy VV~29c lies in the background.
}
  \label{sketch.fig}
\end{figure}

The WSRT observations reveal that the VV~29 system comprises at least three
substantial components, a, b, and the newly identified VV~29c. Close
inspection of deep optical images, in combination with clues from
the 21cm maps, reveals a still more detailed picture, which is
summarized in the sketch in Fig.~\ref{sketch.fig}. Figure~\ref{deep_opt.fig}
shows a filtered, smoothed $g$-band INT image from Trentham et al (2001),
adjusted to high contrast to accentuate the dim features. The image
shows a very low-level ``counter-arm'' to the west extending
to the location of the  HI detection of the `west plume' in
Figs.~\ref{NHI_GALAXY.fig} and \ref{RA_V_MAJ.fig}.  The deep image
also has a clear vertical plume rising northward from the eastern side of
VV~29a, corresponding to the distorted HI contours in Fig.~\ref{NHI_GALAXY.fig};
comparison with lower contrast optical images indicates that the vertical
plume attaches smoothly to a broad arc of faint optical emission to the
south of VV~29a, which we interpret as a sort of partial polar ring or
equivalent to the Magellanic Stream of the Local Group. Using $u$-, $g$-, $r$-,
and $i$-band INT images from Trentham et al (2001), photometry measurements of several
regions of the system reveal that
the compact optical region at the position of VV29c and the long, thin central
part of the plume are bluest. Several blue knots can be identified in the region
of the plume, most likely corresponding to recently star forming regions. The compact
blue source at the position of VV29c is embedded or, more likely, partly obscured by
a significantly redder region that seems to be part of VV29a.

From the radio and optical data, it appears most likely that
the newly identified VV~29c
is located behind the host VV~29a and is receding from it, after having
suffered a near collision with VV~29a.
Any doubts might be resolved by optical and ultraviolet observations
that would sense the dust extinction caused by the spiral arms of the
VV~29a against VV~29c, if VV~29c is in the background.

The tidal plume VV~29b to the east
displays the morphology and kinematics that Toomre and Toomre (1972)
described with restricted N-body calculations three decades ago.  The gas
appears to be both kinematically and morphologically connected to  gas
on the western side of the host, although
the extreme western cloud/plume appears to represent the tip of
a `counter-tail,' stimulated in outskirts of VV~29a
by the interaction.  The bulk of the gas and the
visible starlight in the long filament VV~29b are concentrated in the
eastern plume at a distance of $\sim$40 to 80$h^{-1}$~kpc from the host.

Both Toomre \& Toomre (1972)
and later Wallin and Stuart (1992) examined a range of the collision parameters
that specify the orientation of the axis of rotation relative to the orientation
of orbital angular momentum of the interaction.  The greatest disruption is
inflicted on galaxies whose rotational angular momentum vector aligns with the
orbital angular momentum vector.  In the VV~29 case, we visualize the most massive
galaxy,  the VV~29a host that we view nearly edge-on, as the target galaxy.
A fully intact, late-type, less-massive intruder is incident on the target,
following a trajectory that
carries it by VV~29a approximately in its disk plane, passing with the
same sense of revolution
as the target. This type of trajectory would
be the most effective at raising a tail and counter-tail from the outskirts
of the massive target VV29a.
What makes VV~29 special in relation to simulated and other observed systems,
like for example NGC 7252 (Dubinsky, Mihos \& Hernquist 1996, Hibbard et al. 1994),
is the compactness of the secondary galaxy and the high relative velocity.

Since the time of close encounter,
VV~29c is likely to have traveled a distance that is similar to the distance
traveled by the
outer gas clouds in the VV~29b plume, so we estimate
that it now lies roughly 100~kpc behind VV~29a.  The further implication is
that from our viewpoint VV~29c passed from west to east
in front of VV~29c and experienced
an interaction that diverted its lateral motion to largely recessional
motion with a slight westerly component to carry it back behind
VV~29a.  In this picture, we would expect VV~29c to eventually emerge from
behind VV~29a on the western side.
The interaction slings the VV~29b plume eastward and leaving
it with a slight
recessional velocity component with respect to the target VV~29a.

A plausible case can be made that we are viewing the product of
tidal interaction between two normal galaxies.
The dynamical components of the interaction are detectable through their
baryon content -- stars and hydrogen gas.

\begin{acknowledgements} 
  The authors are grateful to R. Sancisi for valuable comments and
  to the staff of the Westerbork Telescope
  for their skill and dedication in bringing the upgraded
  Multi-Frequency Front Ends and the DZB correlation spectrometer
  systems into operation.
The Westerbork Synthesis Radio Telescope (WSRT) is operated by the
Netherlands Foundation for Research in Astronomy (NFRA) with financial
support of the Netherlands Organization for Scientific Research (NWO).
The deep optical images were
made publically available through the Isaac Newton Groups' Wide
Field Camera Survey Programme. The Isaac Newton Telescope is operated on the
island of La Palma by the Isaac Newton Group in the Spanish Observatorio del
Roque de los Muchachos of the Instituto de Astrofisica de Canarias.
This research has made use of the NASA Astrophysics Data
  System (ADS) and the NASA/IPAC Extragalactic Database
  (NED) which is operated by the Jet Propulsion Laboratory, California
  Institute of Technology, under contract with the National
  Aeronautics and Space Administration. 

\end{acknowledgements}

\end{document}